\documentclass[letterpaper,twocolumn,longbibliography,accepted=2023-03-14]{quantumarticle}
\pdfoutput=1

\usepackage[numbers, sort&compress]{natbib}
\usepackage{amsthm, amsmath, amssymb}
\usepackage{tensor, braket, physics, siunitx} 

\usepackage{graphicx, placeins}
\usepackage[caption=false]{subfig}
\usepackage{float}

\usepackage{hyperref, xcolor}
\usepackage[nameinlink,capitalize]{cleveref}

\usepackage{orcidlink}
\graphicspath{ {./images/} }

\begin{document}
\title{When does a particle arrive?}
\author{Simone Roncallo\,\orcidlink{0000-0003-3506-9027}}
	\email{simone.roncallo01@ateneopv.it}
	\affiliation{Dipartimento di Fisica, Università degli Studi di Pavia, Via Agostino Bassi 6, I-27100, Pavia, Italy}
	\affiliation{INFN Sezione di Pavia, Via Agostino Bassi 6, I-27100, Pavia, Italy}
	
\author{Krzysztof Sacha\,\orcidlink{0000-0001-6463-0659}}
	\email{krzysztof.sacha@uj.edu.pl}
	\affiliation{Instytut Fizyki imienia Mariana Smoluchowskiego, Uniwersytet Jagiello\'{n}ski, ulica Profesora Stanisława Łojasiewicza 11, PL-30-348 Krak\'{o}w, Poland}
	
\author{Lorenzo Maccone\,\orcidlink{0000-0002-6729-5312}}
	\email{lorenzo.maccone@unipv.it}
	\affiliation{Dipartimento di Fisica, Università degli Studi di Pavia, Via Agostino Bassi 6, I-27100, Pavia, Italy}
	\affiliation{INFN Sezione di Pavia, Via Agostino Bassi 6, I-27100, Pavia, Italy}

\begin{abstract}
    We compare the proposals that have appeared in the literature to describe a measurement of the time of arrival of a quantum particle at a detector. We show that there are multiple regimes where different proposals give inequivalent, experimentally discriminable, predictions. This analysis paves the way for future experimental tests.
\end{abstract}
\maketitle

\section*{Introduction}
The trajectory of a classical system can be uniquely determined from its dynamics and initial conditions. We can invert it to compute the value of the time at which the object is at a certain position. If this is the position of the detector, the time of detection is commonly known as time of flight or time of arrival (TOA).

Determining the TOA of a quantum system is far more complicated. In quantum mechanics each observable quantity must be described by a self-adjoint operator that acts on the Hilbert space of the system, or by a positive-operator valued measure. Usually the quantization of a classical system is done by means of the correspondence principle, i.e. by promoting each classical observable to a self-adjoint operator on the Hilbert space. This procedure cannot be directly employed for time, which is a (scalar) parameter in textbook quantum mechanics. There are several reasons for this, e.g. the Pauli objection \citep{book:Pauli}: since the energy is responsible for time translations, any time operator must be conjugated to an energy operator whose spectrum then, must be continuous and unbounded. However, the Hamiltonian does not usually satisfy such a condition, preventing the possibility to directly construct a self-adjoint time operator.
\begin{figure}[H]
	\centering
	\includegraphics[width = 0.95\linewidth]{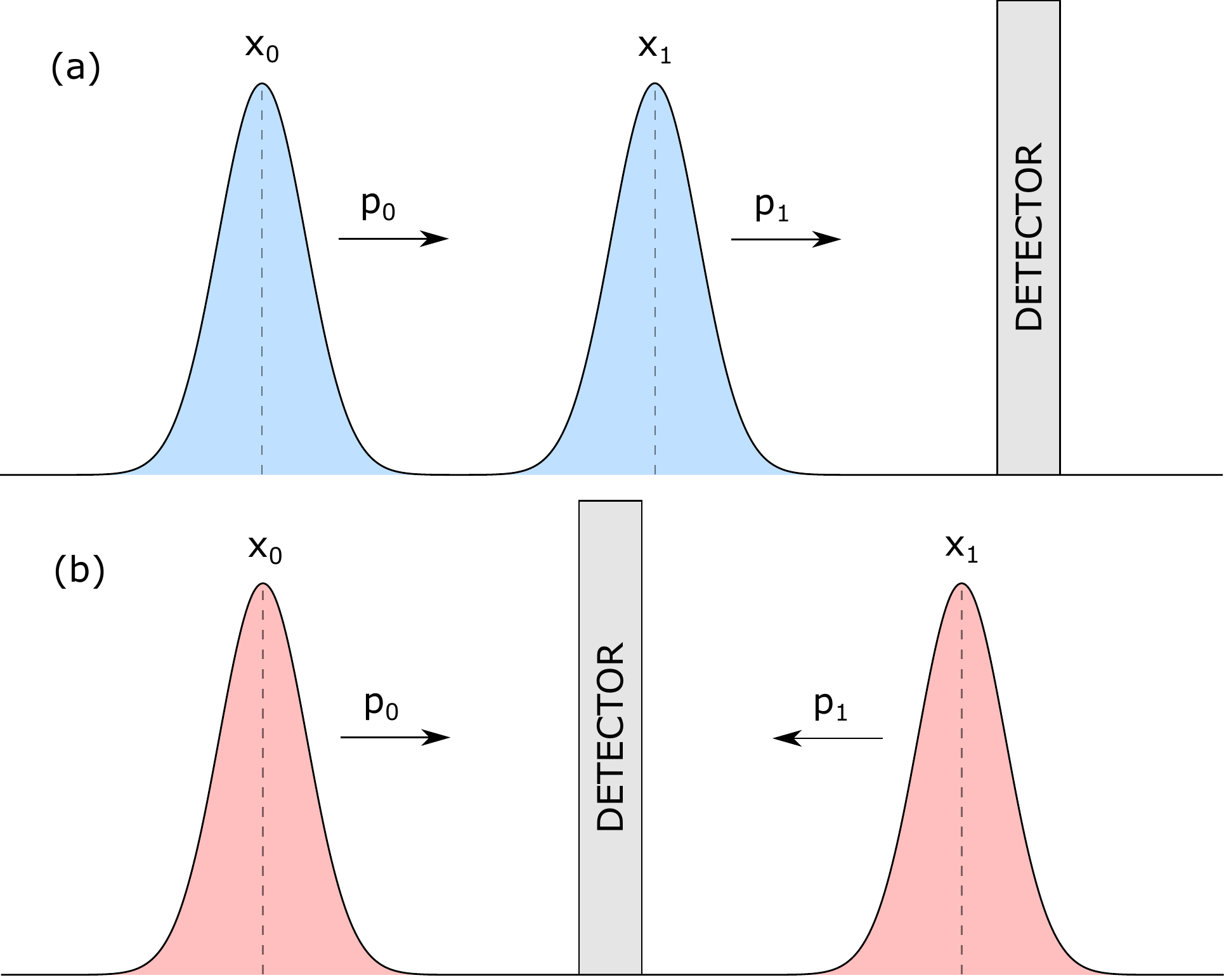}
    \caption{Qualitative representation of the regimes considered in our comparison. A particle is initially prepared in a superposition of two Gaussian wave packets. When the packets overlap near the detector, TOA interference is observed. (a) In this regime two packets with initial average positions $x_0$, $x_1$ cross the detector from the same side, with average momenta $p_0$, $p_1$. A limit case is provided by the \emph{overtaking condition}, in which the packets completely overlap at the detector with the same average TOA. (b) In this scenario two wave packets cross the detector from opposite directions. We refer to this regime as \emph{counter-propagating}.\label{fig:setup}}
\end{figure}

In experiments the TOA is usually computed using the semiclassical approximation \citep{book:Frohlich-Durr,book:Feynman_PathIntegral}, which holds only in regimes where quantum effects can be neglected, i.e. if the particle is approximated by a well localized wave packet, with negligible spatial uncertainty compared to the detector dimension \citep{art:Das}. Due to the limited validity of this approach, several quantum time and quantum TOA formulations have been proposed. Some of these have been achieved in the framework of standard quantum mechanics, e.g. through canonical quantization \citep{art:Aharonov-Bohm,art:Grot-Rovelli-Tate,art:Galapon} or following different constructions \citep{art:Kijowski,art:Delgado-Muga, book:Werner, art:Werner_Screen, art:Unruh,art:Nikolic}, while others are derived from extensions or alternative formulations of the quantum theory \citep{art:Aharonov_QReferenceFrame, art:Aharonov_NewUniverse, art:Rovelli_RelationalQM, art:Rovelli-Reisenberger, art:Page-Wooters,art:Maccone-Sacha,art:Maccone_QuantumTime,art:Fredenhagen-Brunetti} and also within the framework of Bohmian mechanics \citep{book:Frohlich-Durr,art:Das_Spin,art:Leavens_BohmianTOA,art:ScientificAmerican} (see \citep{book:Muga_Vo1, book:Muga_Vo2} for a comprehensive  selection of works about time in quantum mechanics).

In this paper we review and compare some of these proposals.
We analyse the interference patterns that arises in the TOA probability distribution, when considering a state described by a superposition of Gaussian wave packets. We explore different regimes (see \cref{fig:setup}) where we can compare the Kijowski's axiomatic construction \citep{art:Kijowski} (which is in agreement with \citep{art:Aharonov-Bohm,art:Grot-Rovelli-Tate,art:Galapon}), the quantum flux approach \citep{art:Delgado-Muga,art:Das}, the semiclassical approximation \citep{book:Feynman_PathIntegral} and the quantum clock proposal \citep{art:Maccone-Sacha,art:Maccone_QuantumTime}. Our discussion is not limited to predicting a specific experimental result, nor it specializes to any particular model of particle detection: it highlights those regimes in which different theoretical approaches lead to contradicting predictions, by analysing the distribution of times at which the particle occupies a certain position, i.e. where the detector should be placed. Finally, we explore a possible experimental implementation to discriminate between these TOA proposals, using a superposition of packets obtained by applying Bragg diffraction \citep{art:BraggScattering} to a Bose-Einstein condensate trapped in an accelerator ring \citep{art:VonKlitzing}.

The paper is outlined as follows. In \cref{sec:SectionI} we review the Kijowski's axiomatic construction, the quantum flux approach, the semiclassical approximation and the quantum clock proposal. In \cref{sec:SectionII} we compare them for different superpositions of Gaussian wave packets. In \cref{sec:SectionIII} we analyse the previously mentioned experimental regime.

\section{Time of arrival distributions\label{sec:SectionI}}
In this section we review some of the proposed TOA distributions.

\subsection{The Kijowski's proposal\label{sec:SectionIA}}
A potential solution to the TOA problem is discussed by Kijowski \citep{art:Kijowski}, adopting an axiomatic approach. In principle, the Kijowski's axioms can be potentially satisfied by infinitely many candidates of a TOA distribution $\Pi$. However, there exists a \emph{unique} choice of $\Pi$ which best approximates any other possible candidate $\Pi'$, by providing the same TOA expectation value of any $\Pi'$ and a lower bound to its second moment. For simplicity, we restrict the results of \citep{art:Kijowski} to the case of a one-dimensional particle, for a detector placed in the origin.

We consider a particle with either positive or negative momenta $p$, i.e. described by a wave function $\Psi$ with momentum representation supported for either positive or negative values of $p$. Respectively, two probability distributions $\Pi_+$, $\Pi_-$ are obtained
\begin{equation}
    \Pi_{\pm}(t) = \frac{1}{4\pi^2 m \sqrt{\hbar}}\left| \ \int_{0}^{\pm \infty} dp \ \widetilde{\Psi}(p,t) \sqrt{|p|} \ \right|^2 \ , 
    \label{eq:Kijowski}
\end{equation}
with $m$ the mass of the particle and
\begin{equation}
    \widetilde{\Psi}(p,t) = \exp\left(-\frac{itp^2}{2m\hbar}\right) \tilde{\Psi}(p) \ .
\end{equation}
Here $\widetilde{\Psi}(p) = \langle p|\Psi(0)\rangle$ denotes the Fourier transform of the wave function at $t=0$. The distribution of Kijowski can be rewritten using the wave function in the position representation. This equivalent formulation is due to Leavens \citep{art:Leavens_Kijowski}, which reads
\begin{equation}
	\Pi_{\pm}(t) = \frac{\hbar}{32\pi m}\left| \ \int^{+\infty}_{-\infty} dx \ \frac{1\pm i \text{sign}(x)}{|x|^{3/2}}g(x,t) \ \right|^2 \ ,
	\label{eq:Leavens}
\end{equation}
with\begin{equation}
	g(x,t) = \left[\Psi(x,t) - \Psi(0,t) \right] \ . 
\end{equation}
Regardless of the representation, for a generic particle whose momentum is not supported only on positive or negative $p$, the two contributions are combined so that the total TOA probability distribution reads (see e.g. \citep{art:Leavens_Paradoxes} and references therein)
\begin{equation}
	\Pi_{K}(t) =  \frac{\Pi_+(t) +  \Pi_-(t)}{N_{K}} \ ,
	\label{eq:TotalKijowskiSum}
\end{equation}
with the normalization constant $N_{K}$ given by
\begin{equation}
    N_{K} =  \int_{-\infty}^{+\infty}dt \sum_{\alpha \in \{+,-\}} \Pi_{\alpha}(t) \ .
    \label{eq:total_normalization}
\end{equation}
The results of Kijowski reproduce the distribution that Aharonov and Bohm obtained through the quantization of the classical TOA, based on the correspondence principle \citep{art:Aharonov-Bohm}. See \citep{art:Das_Gauge} for a review and a comparison between these two results. 

Other proposals reproduce or are at least in agreement with the predictions of Kijowski. For example, in \citep{art:Grot-Rovelli-Tate} the quantization of the classical TOA uses a different symmetric ordering of the operator, with respect to the approach of Aharonov and Bohm. When considering a particle with either positive or negative momentum, this TOA distribution takes the same form of \cref{eq:Kijowski}.

Instead of a free particle, a different solution consists in confining the motion within a given space interval. This idea has been followed by Galapon, who considers the discrete probability distribution of a particle quantized in a finite box \citep{art:Galapon}. In the limit of an infinite interval, this reproduces the same distribution of \cref{eq:Kijowski}.

In \citep{art:Leavens_Paradoxes} Leavens discusses some paradoxical behaviours arising from the Kijowski's TOA: for a particle that propagates towards an infinite potential barrier, the distribution predicts non-vanishing probabilities even in prohibited regions. In this regard, further comments are given in \citep{art:Muga-CommentLeavens,art:Leavens-ReplyMuga} while other paradoxes and issues are analysed in \citep{art:Das_Gauge}.

\subsection{The quantum flux proposal\label{sec:SectionIB}}
A straightforward way to obtain a TOA distribution starts from the definition of the Schrödinger current operator, or quantum flux, as
\begin{equation}
	\hat{J} = \frac{\hat{p}}{2m}\ket{x}\bra{x} + \frac{1}{2m}\ket{x}\bra{x}\hat{p} \ ,
	\label{eq:CurrOperator}
\end{equation}
with $m$ the mass of the particle, $x$ the eigenvalues of the position operator and $\hat{p}$ the momentum operator. By taking the expectation value of \cref{eq:CurrOperator} on the state vector $\ket{\Psi(t)}$, it yields
\begin{equation}
	\Pi_{F}(x,t) = \frac{\hbar}{N_{F} m}\Im[\Psi^*(x,t)\partial_x \Psi(x,t)] \ ,
	\label{eq:QuantumFlux}
\end{equation}
with normalization constant $N_{F}$ given by \citep{art:Delgado-Muga}
\begin{equation}
    N_{F} = \int_{-\infty}^{+\infty}dt \ \Pi_{F}(t) \ .
    \label{eq:QF_normalization}
\end{equation}
For a detector placed at the origin we denote the TOA distribution as $\Pi_{F}(t) = \Pi_{F}(0,t)$.

In \citep{art:Kijowski} Kijowski notes that such an operator cannot be generally interpreted as a probability current, due to lack of a condition that makes \cref{eq:CurrOperator} positive-definite. In \citep{art:Delgado-Muga} Delgado and Muga shows that this interpretation is possible under a specific choice of states. To this extent, they construct a self-adjoint operator whose orthonormal and complete set of eigenstates are well-behaved under time translations, while being conjugated not to the Hamiltonian of the system but to another operator with dimensions of energy. Such energy-like operator is the Hamiltonian multiplied by the sign of the momentum operator. They show that \cref{eq:CurrOperator} is positive-definite only for states with definite momentum sign, thus admitting a probabilistic interpretation. The TOA distribution is then obtained by projecting such states along the basis of this self-adjoint temporal-like operator. 

The TOA probability distribution of \cref{eq:QuantumFlux} can also be derived in the framework of Bohmian mechanics \citep{art:Das, art:Das_Gauge,book:Frohlich-Durr}. In \citep{book:Frohlich-Durr} the authors discuss the existence of free particle states which forbid the probabilistic interpretation of the quantum flux, leading to negative probability contributions even for definite momenta signs. This eventuality takes the name of \emph{quantum backflow}, which was historically introduced by Bracken and Melloy \citep{art:Bracken}. We will show that even in presence of backflow in the quantum flux, the quantum clock and Kijowski's TOA distributions take only positive values.

\subsection{The semiclassical approximation\label{sec:SectionIC}}
Experimentally, time measurements are often treated semiclassically as a momentum measurement \citep{book:Frohlich-Durr,book:Feynman_PathIntegral}, from which the TOA distribution reads
\begin{equation}
 	\Pi_{sc}(x,t) = \frac{mL}{t^2}\left| \widetilde{\Psi}\left( \frac{mL}{t} \right) \right|^2 \ , 
 	\label{eq:Semiclassical}
\end{equation}
with $\widetilde{\Psi}(p)$ the components of the Fourier transform of $\Psi(x,0)$ and $L$ the distance between the source of the wave and the detector \citep{book:Frohlich-Durr,art:Das}. Due to the nature of this definition, \cref{eq:Semiclassical} gives sensible results only for those states that possess a trajectory interpretation (e.g. Gaussian wave packets), and for which the width of the wave function at $t=0$ can be considered negligible with respect to the source-detector distance. 

\subsection{The quantum clock proposal\label{sec:SectionID}}
So far we have considered only proposals that treat time as a label attached to the particle. There are alternative points of view in which time arises as a property of a quantum reference frame: the \emph{quantum clock} of a physical system \citep{art:Page-Wooters,art:Aharonov_QReferenceFrame}. An example is given by the Page and Wooters mechanism, in which time and dynamics emerge from the entanglement between the system and the clock \citep{art:Page-Wooters}. Several criticisms have been addressed to this (e.g. see \citep{art:Kuchar}). To overcome them, a modification of the mechanism has been recently discussed \citep{art:Maccone_QuantumTime}. This slight extension correctly reproduces the standard quantum mechanics predictions such as measurements statistics and propagators, under proper conditioning of the entangled state. In this framework a well-defined time operator  can be constructed, giving rise to time-energy uncertainty relation, while bypassing the above mentioned Pauli objection \citep{art:Maccone-Leon}. 

As shown in \citep{art:Maccone-Sacha}, we can employ the quantum clock approach to describe generic quantum time measurements, like the TOA of a particle at the detector. We now review the construction of the quantum clock TOA. The Hilbert space of the physical system $\mathcal{H}_{S}$ is equipped with an ancillary one $\mathcal{H}_{C}$, that describes time as ``what is measured by the clock'', with a Hamiltonian linear in momentum. This last condition is necessary to obtain a Schrödinger dynamics. The global Hilbert space is constructed as $\mathcal{H} = \mathcal{H}_{S} \otimes \mathcal{H}_{C}$, so do the global states $\ket{\Phi}$ that are the eigenvectors of the total Hamiltonian, under a constraint with the form of the Wheeler–DeWitt equation \citep{art:DeWitt}. These states are static, simultaneously encoding the history of the system in the time interval $\mathcal{T} = [-T/2,T/2]$
\begin{equation}
	\ket{\Phi} = \frac{1}{\sqrt{T}} \int_{\mathcal{T}} dt \ket{t} \otimes \ket{\Psi(t)} \ ,
	\label{eq:global_state}
\end{equation}
where $T \to \infty$ is a regularization parameter. Here $\ket{\Psi(t)}$ is the time-dependent state so far considered, which is obtained from \cref{eq:global_state} by conditioning on $t$, i.e. projecting $\ket{\Phi}$ on $\ket{t}$. 

For a dimensionless detector placed at $x$, a positive-operator-valued measure (POVM) describes a joint measurements on the particle and on the clock as
\begin{align}
	&\forall t : \Theta_t = \ket{t}\bra{t} \otimes \ket{x}\bra{x} \ , \\
	&\Theta_{na} =  1 - \int dt \ \Theta_t \ ,
\end{align}
where each value of $t$ labels a different element of the POVM \citep{art:Maccone-Sacha}. Using the Born rule together with the Bayes formula on $\Theta_t$, for $x=0$ the TOA distribution reads
\begin{equation}
	\Pi_{C}(t) = \frac{1}{N_{C}(T)}|\Psi(0,t)|^2 \ ,
	\label{eq:qc-distribution}
\end{equation}
that is the conditioned probability that the particles arrives at the detector with time of flight $t$, with normalization constant $N_{C}(T)$ that depends from the regularization parameter $T$ by
\begin{equation}
	N_{C}(T) = \int_{-T/2}^{+T/2} dt \ |\Psi(0,t)|^2 \ .
	\label{eq:QC_normalization}
\end{equation}
The TOA distribution \eqref{eq:qc-distribution} becomes independent of $T$ if the probability that the particle remains at the detector at arbitrary large times is sufficiently low, namely whenever the integral in \cref{eq:QC_normalization} converge in the limit $T \to + \infty$.

Similar results are also obtained by Fredenhagen and Brunetti in \citep{art:Fredenhagen-Brunetti}, where they discuss a generalization of quantum states in term of particle weights \citep{art:Porrmann_Weights} and in a way that is consistent with the Gelfand–Naimark–Segal (GNS) representation theorem. In \citep{art:Gambini}, Gambini and Pullin combine the Page and Wooters mechanism with Rovelli's evolving constant of motions, both for a clock Hamiltonian that is unbounded or bounded-below. In the former case, i.e. when the clock Hamiltonian is linear in momentum, this yields the same TOA distribution of \cref{eq:QC_normalization} normalized on the whole time interval. The authors discuss that, in contrast to the Kijowski's proposal, no paradoxical behaviour occurs for a particle travelling towards an infinite potential barrier, since $\Pi_C$ predicts no TOA in the prohibited regions. 
 
Before discussing our comparisons, we address the normalizability of each TOA distribution. For simplicity we consider a Gaussian wave packet (see \cref{eq:gaussian_SI} below). At later times the behaviour of the wave packet can be approximated as $|\psi(0,t)|^2 \simeq e^{-p_0^2}/t$. Such contribution diverges logarithmically when integrated on the whole time domain, yielding 
\begin{equation}
	N_{C}(T) = \int_{-T/2}^{+T/2} dt \ |\psi(0,t)|^2 \simeq 2e^{-p_0^2} \log \frac{T}{2} \ . 
	\label{eq:divergence}
\end{equation}	
As a consequence $\Pi_{C}$ turns out to be non-normalizable. 

Such an obstruction does not appear neither in the Kijowski nor in the quantum flux proposals. In the former, the normalization is guaranteed a priori and independently from the choice of the packet (see the axiomatic construction of Kijowski in \citep{art:Kijowski}). In the latter, a Gaussian wave packet gives $\partial_x \psi(x,t) = \psi(x,t)(-x + x_0 + ip_0)/(1+it)$. In the same regime this yields $\Im[\Psi^*(0,t)\partial_x \Psi(0,t)] \simeq p_0/t^3$, which implies that the normalization constant converges on the whole time interval.
\begin{figure}[t]
    	\centering
    	\includegraphics[width = \linewidth]{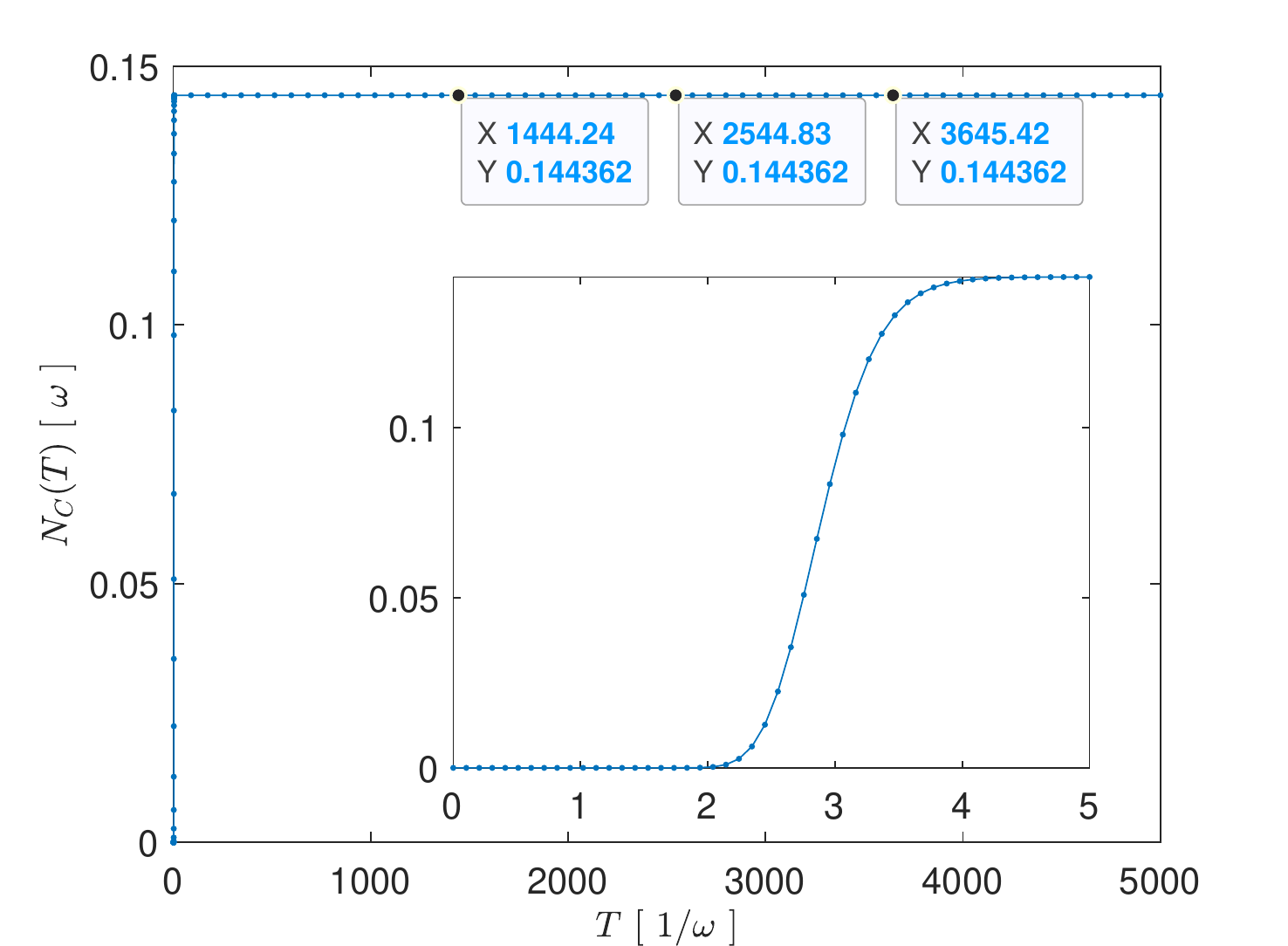}
    	\caption{Normalization constant of the quantum clock TOA for a single right-moving Gaussian wave packet, see \cref{eq:gaussian_SI}. The parameters of the packet are $x_0 = -10 \ l_0$, $p_0 = 7 \ \hbar/l_0$ and $\sigma_0 = 1 \ l_0$. We observe an initial increase of $N_{C}$ as $T \lesssim 5 \ t_0$, that is when the bulk of the wave packets approaches the detector. For greater values of the regularization parameter the integral shows a slow logarithmic growth which is not appreciable in this graph, produced by the trailing tale of the packet that spreads while crossing the detector. For a reasonable choice of the integration domain, no numerical difference arises in $N_{C}$.\label{fig:normalization}}
\end{figure}

The behaviour of $\Pi_{C}$ is due to the choice of a Gaussian wave packet, which is, in turn, non-normalizable. Unlike the other proposals, the quantum clock directly depends from the probability density of the wave function, thus it inherits its properties, including the non-normalizability. We observe that such an issue arises also classically. For example, let us consider the motion of a classical cloud that propagates towards a detector from the left, while spreading due to diffusion. We can describe the probability density of the cloud using $|\psi|^2$ from \cref{eq:gaussian_density_SI}. When the diffusion is non-negligible with respect to the global translation of the cloud, its left tail may spread faster than the way it is approaching the detector. This is what happens also in \cref{eq:divergence}, where a faster translation of the packet (controlled by $p_0$) results in a slower divergence of the normalization constant. In this regime, it is however impossible to completely detect the cloud even by keeping the detector switched on indefinitely. This results in a non-normalizability of the TOA distribution. 

These considerations suggest that it is ill advised to
postulate that the TOA distribution should be normalized as is done in Kijowski \citep{art:Kijowski}. Indeed, there are trivial situations in which it cannot be normalizable at all, such as when the particle is stationary at the detector position or in the above scenario, when the particle wave function spreads so quickly that the trailing tail has sufficiently large probability to be found at the detector position at arbitrarily large times.

In practical situations this is not an issue because typically the divergence is not appreciable (see \cref{fig:normalization}). In any case, the value of the regularization parameter $T$ is given by the total duration of the experiment. For our purpose, in the next section we consider $T = 100 \ 1/\omega$ (in the units introduced below).

\section{Comparisons\label{sec:SectionII}}
We now present a comparison of the previous TOA proposals in several scenarios. We focus on the TOA interference patterns of Gaussian wave packets that overlap near a detector, whose cross-sectional area is much greater than the transversal width of each packet, so that a click can happen whenever a particle reaches the longitudinal position of the detector. In this section, we perform our comparisons by analysing the distribution of times at which the particle occupies a certain position, without specializing to any particular detection model, i.e. by keeping the TOA independent of any particle-detector interaction. Moreover, we restrict to those regimes in which each packet can be detected at most once, so that its subsequent collapse provides no contributions in the TOA distribution.

Our results highlight the differences between the previously introduced proposals, allowing to discriminate between them.  We keep $\Pi_{K}$ and $\Pi_{F}$ separate, underlining those regimes in which they produce significantly different predictions, e.g. when quantum backflow contributions cannot be neglected.

We consider a Gaussian wave packet with initial position $x_0$, momentum $p_0$ and standard deviation $\sigma_0$
\begin{equation}
    \psi(x,0) = \frac{1}{(\pi \sigma_0^2)^{1/4}}e^{-(x-x_0)^2/2\sigma_0^2 + ip_0(x-x_0)/\hbar} \ .
\end{equation}
Its time evolution, governed by the free Hamiltonian $H = \hat{p}^2/2m$, reads
\begin{equation}
    \psi(x,t) = A(t)e^{\frac{-(x-x_0-p_0t/m)^2}{2\sigma_0^2(1 + i \hbar t/m\sigma_0^2)}}e^{\frac{ip_0}{\hbar}\left(x-x_0-\frac{p_0t}{2m}\right)}
    \label{eq:gaussian_SI}
\end{equation}
with
\begin{equation}
    A(t) = \left[\pi^{1/2}\left(\sigma_0 + \frac{i\hbar t}{m\sigma_0}\right)\right]^{-1/2} \ .
\end{equation}
The probability density is
\begin{equation}
    |\psi(x,t)|^2 = \frac{e^{\frac{-(x-x_0-p_0t/m)^2}{\sigma_0^2 + \hbar^2t^2/m^2 \sigma_0^2}}}{\sqrt{\pi}\sqrt{\sigma_0^2 + \hbar^2t^2/m^2 \sigma_0^2}} \ .
    \label{eq:gaussian_density_SI}
\end{equation}
Unless explicitly stated, we work in units of length, time and energy respectively given by $l_0 = \sqrt{\hbar/m\omega}$, $t_0 = 1/\omega$ and $E_0 = \hbar \omega$, with $\omega$ the frequency of the harmonic trap where the particle is initially prepared in the ground state. After the preparation, the trap is turned off and the particle is kicked so that its probability density is described by \cref{eq:gaussian_density_SI}.
\begin{figure}[t]
    \centering
    \includegraphics[width = \linewidth]{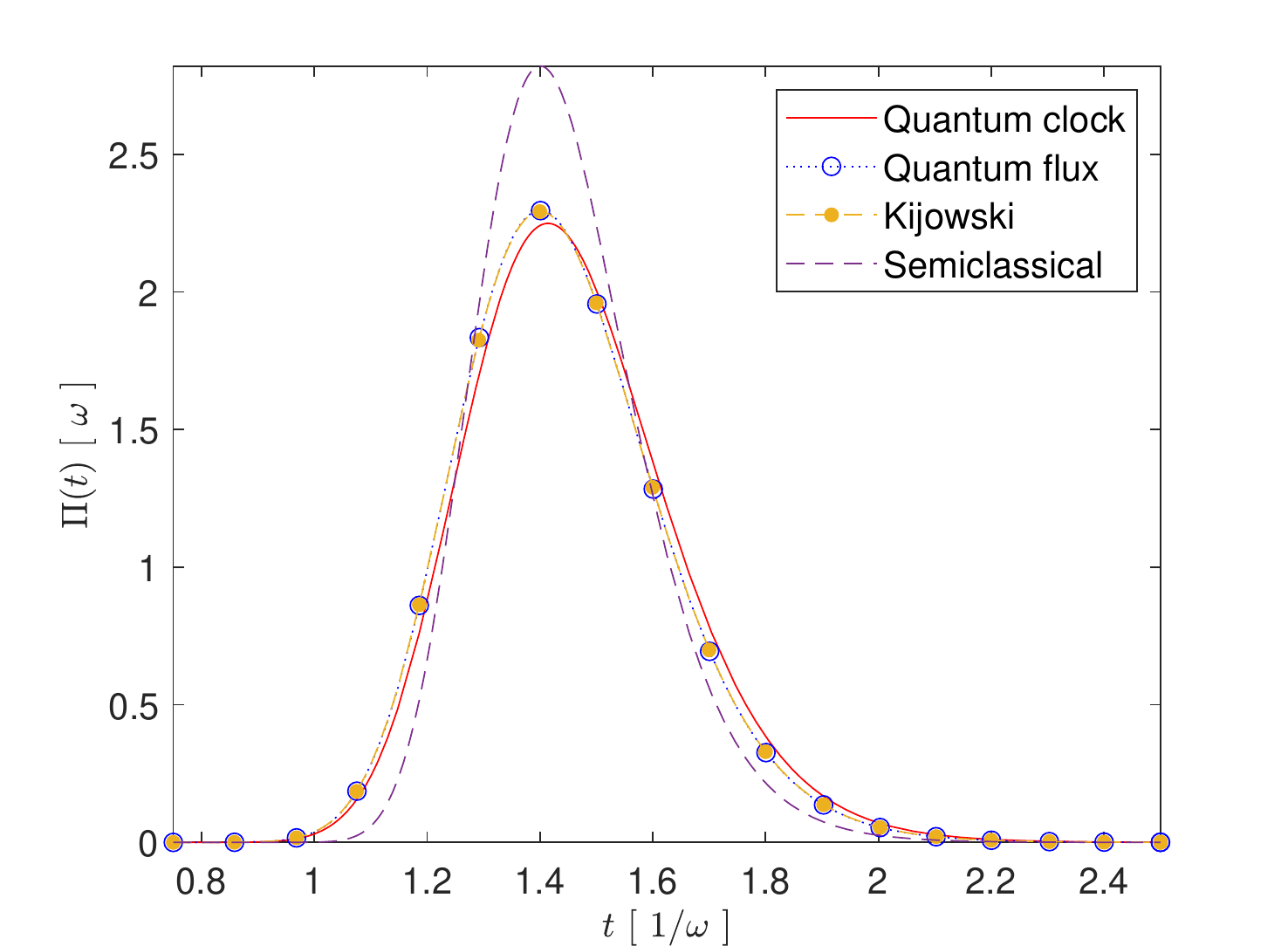}
    \caption{Time of arrival of a single right-moving Gaussian wave packet propagating towards the detector. The parameters of the packet are $x_0 = -10 \ l_0$, $p_0 = 7 \ \hbar/l_0$ and $\sigma_0 = 1 \ l_0$. In this regime $\Pi_{F}$ and $\Pi_{K}$ overlap, while $\Pi_{C}$ and $\Pi_{sc}$ give different predictions.\label{fig:single}}
\end{figure}

\subsection{Gaussian wave packet}
We preliminarily consider the TOA distributions of a single Gaussian wave packet that travels towards the detector from the left, i.e. right-moving, and whose time evolution is given by \cref{eq:gaussian_SI}.

Since Gaussian states allow a trajectory interpretation, in this section we can safely employ also the semiclassical approximation.

From \cref{fig:single} we notice that $\Pi_{C}$ gives a similar prediction with respect to $\Pi_{F}$ and $\Pi_{K}$. The last two overlaps in this regime.
\begin{figure}[t]
    	\centering
    	\includegraphics[width = \linewidth]{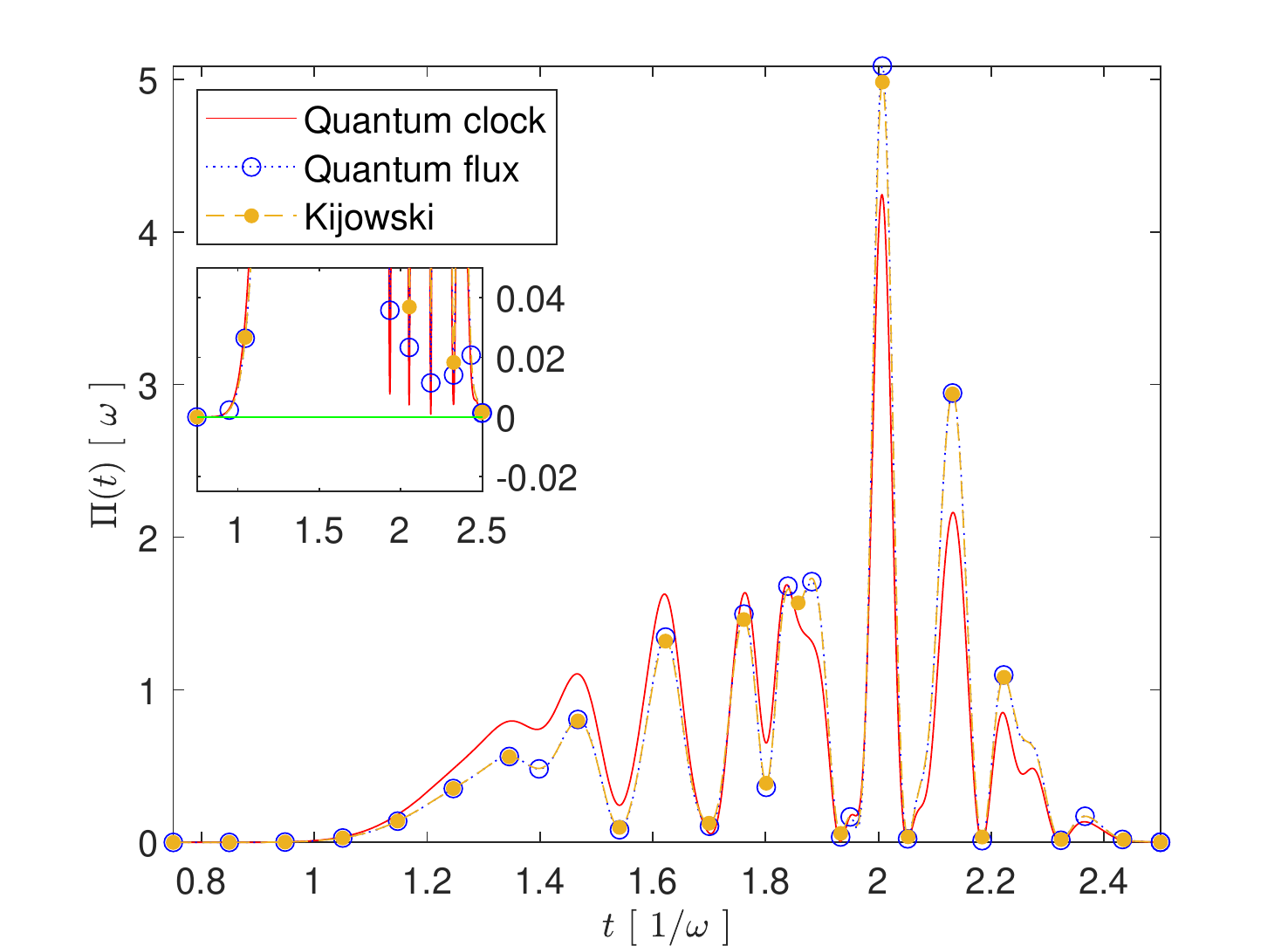}
    	\caption{Time of arrival of a train of $n=4$ right-moving Gaussian wave packets that overlap near the detector. The packets parameters are $x_k = (-10 - 8k) \ l_0$, $p_k = (7 + 3k) \ \hbar/l_0$ and $\sigma_k = 1 \ l_0$, with $k \in \{0,1,2,3\}$. In this regime $\Pi_{F}$ and $\Pi_{K}$ almost overlap (missing markers are due to sampling differences between $\Pi_{K}$ and $\Pi_{F}$), while $\Pi_{C}$ produces significantly different results. Here no negative values occur for the quantum flux, which means that in this regime all the TOA distributions admit a probabilistic interpretation. The inset is a zoom-in of the main plot close to the horizontal axis.\label{fig:overlapping}}
\end{figure}

\subsection{Gaussian wave train and time of arrival interference}
In this section we consider sequences of $n$ Gaussian wave packets, described by the superposition
\begin{equation}
    \ket{\Psi(t)} = \frac{1}{\sqrt{n}}\sum_{k=0}^{n-1} \ket{\Psi_k(t)} \ ,
    \label{eq:Gaussian-train}
\end{equation}
where $\ket{\Psi_k(t)}$ has position representation given by \cref{eq:gaussian_SI}. We denote $x_k$ and $p_k$, respectively, the average position and momentum of the k-th Gaussian wave packet.

An analysis similar to \citep{art:Das} revealed that all the TOA proposals are well-behaved for a Gaussian train that satisfies a no-spreading condition, namely when quantum diffusion is negligible and the propagation of each packet is approximately a rigid translation.\footnote{The no-spreading condition can be employed in the Kijowski's proposal by using, instead, the Leavens prescription of \cref{eq:Leavens}.} In this regime, each TOA distribution yields an equiprobable sequence of peaks centred around the classical TOA $|x_k m/p_k|$. When diffusion is taken into account, a similar behaviour is obtained with a sequence of peaks that show an overall exponential decay. In our comparisons we do not make such approximation and we always consider Gaussian trains that propagates while spreading. 
\begin{figure}[t]
    	\centering
    	\includegraphics[width = \linewidth]{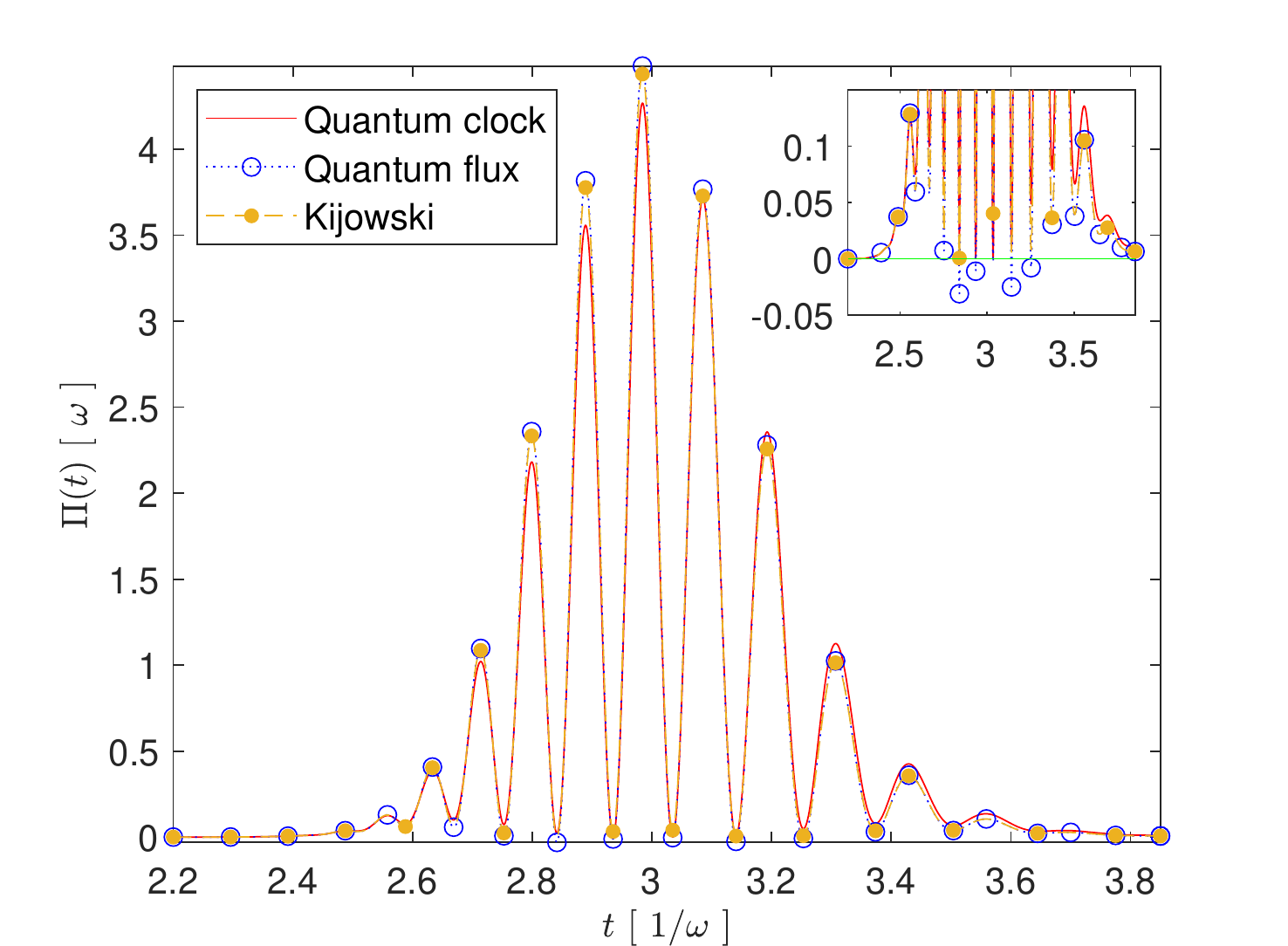}
    	\caption{Time of arrival of $n=2$ right-moving Gaussian wave packets that completely overlap at the detector under the overtaking condition. The parameters of each packet are $x_0 = -30 \ l_0$, $p_0 = 10 \ \hbar/l_0$ and $p_1 = 15 \ \hbar/l_0$ and $\sigma_0 = \sigma_1 = 1 \ l_0$, which give $x_1 = -45 \ l_0$. While the prediction of $\Pi_{F}$ and $\Pi_{K}$ are similar in this regime, $\Pi_{C}$ produces significantly different results. In the inset we show that $\Pi_{F}$ takes negative values due to the presence of quantum backflow, for this reason it cannot be interpreted as a probability distribution (see the text for a detailed discussion).\label{fig:overtaking}}
\end{figure}

\subsubsection{Overlapping packets}
We consider a train of Gaussian wave packets that propagates in the same direction. We adjust the parameters of each packet such that they partially overlap near the detector.

We compute the TOA according to each proposal. In \cref{fig:overlapping} all the distributions show an interference pattern, which is due to the superposition of the packets starting to overlap near $x=0$. However, we observe that $\Pi_{C}$ behaves quite differently from $\Pi_{F}$ and $\Pi_{K}$ in a way that cannot be reabsorbed by a certain choice of the normalization constant.

\subsubsection{Overtaking packets}
We now specialize to the case of two Gaussian wave packets, respectively localized at $x_0$ and $x_1$, with momenta $p_0$ and $p_1$. Given $p_1 > p_0$, we adjust $x_1$ in such a way that both  wave packets arrive at the detector with the same average TOA, that is when the faster but delayed packet overtakes the other one at $x=0$. We refer to this last requirement as an \emph{overtaking condition} which is guaranteed when $x_1/p_1 = x_0/p_0$.

From \cref{fig:overtaking} we see that TOA interference arises with some marked differences between $\Pi_{F}$, $\Pi_{K}$ and $\Pi_{C}$. As happened in the previous regime, this behaviour cannot be justified by a proper redefinition of the normalization factor of $\Pi_{C}$. 

\begin{figure}[t]
    \centering
   	\includegraphics[width = \linewidth]{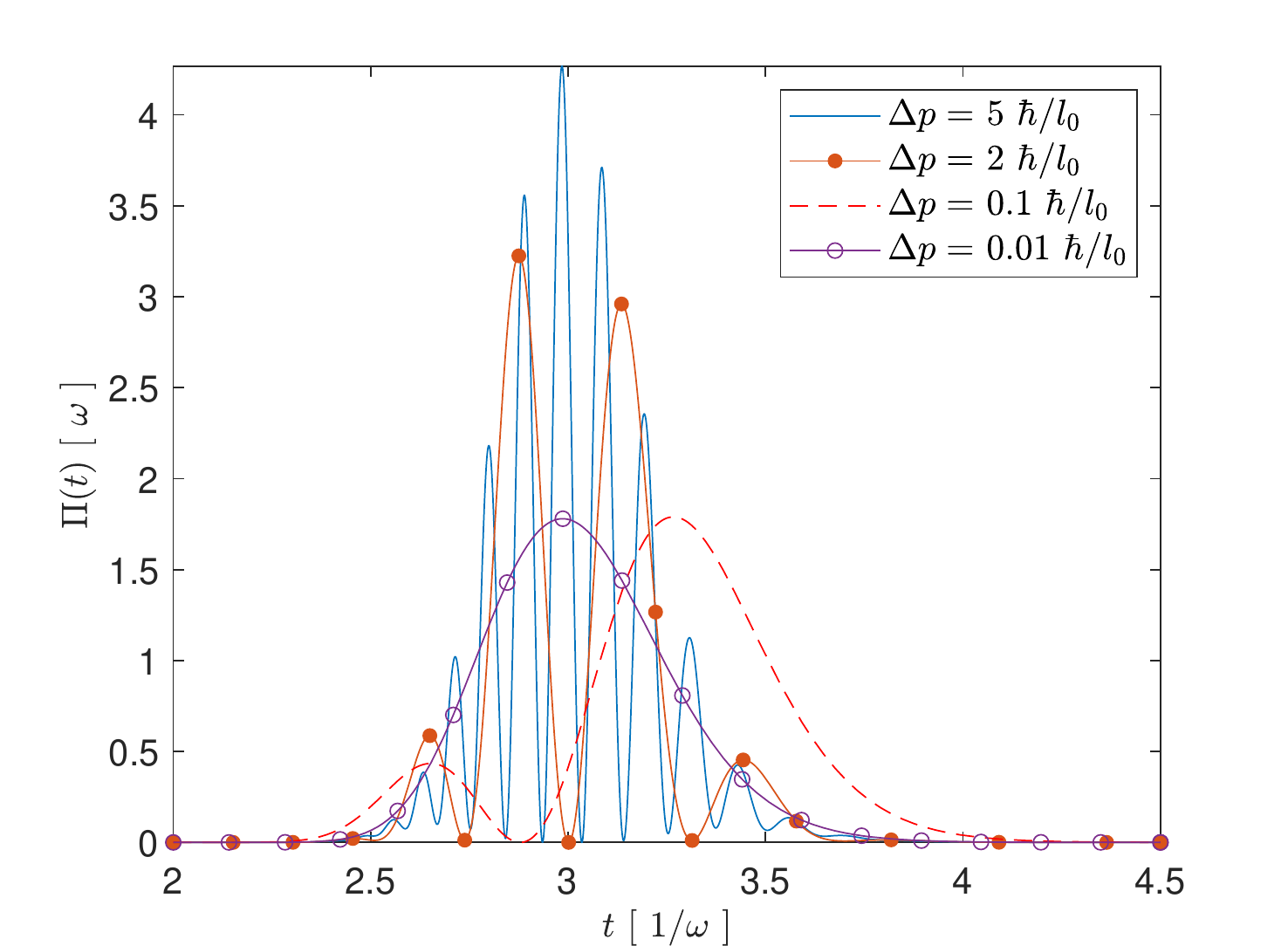}
	\caption{Quantum clock time of arrival of a train of $n=2$ right-moving Gaussian wave packets. We consider different values of the momentum separation $\Delta p$, showing that lower is the value of $\Delta p$, the wider and less numerous are the TOA interference fringes, until a minimum value is reached and interference completely disappears. This behaviour is also reproduced by $\Pi_{K}$ and $\Pi_{F}$\label{fig:momentum-diff}.}
\end{figure}

The number of interference fringes is related to the momentum separation between the two packets $\Delta p = |p_0 - p_1|$. The lower the value of $\Delta p$, the wider and fewer are the fringes in the TOA interference pattern. When this separation is sufficiently low, interference disappears and a result similar to the single packet regime of \cref{fig:single} is obtained. In \cref{fig:momentum-diff} we explicitly show this behaviour for $\Pi_{C}$, that is similarly reproduced by the other TOA proposals.

The quantum flux distribution is not positive-definite in this regime, so its probabilistic interpretation is not allowed here. This issue is due to the presence of quantum backflow \citep{art:Bracken}: although both packets have support for positive momenta only (i.e. they are right-moving), interference may produce a temporary inversion of the motion at the detector. This yields both right and left-moving contributions in the quantum flux current, leading to a TOA distribution with negative probabilities. See the appendix of \citep{book:Frohlich-Durr} for a discussion of backflow in this same regime, but analysed in terms of Bohmian trajectories. 

In this regime, both $\Pi_{K}$ and $\Pi_{C}$ remains positive-definite, although they still produce different results.

\begin{figure}[t]
    \centering
    \includegraphics[width = \linewidth]{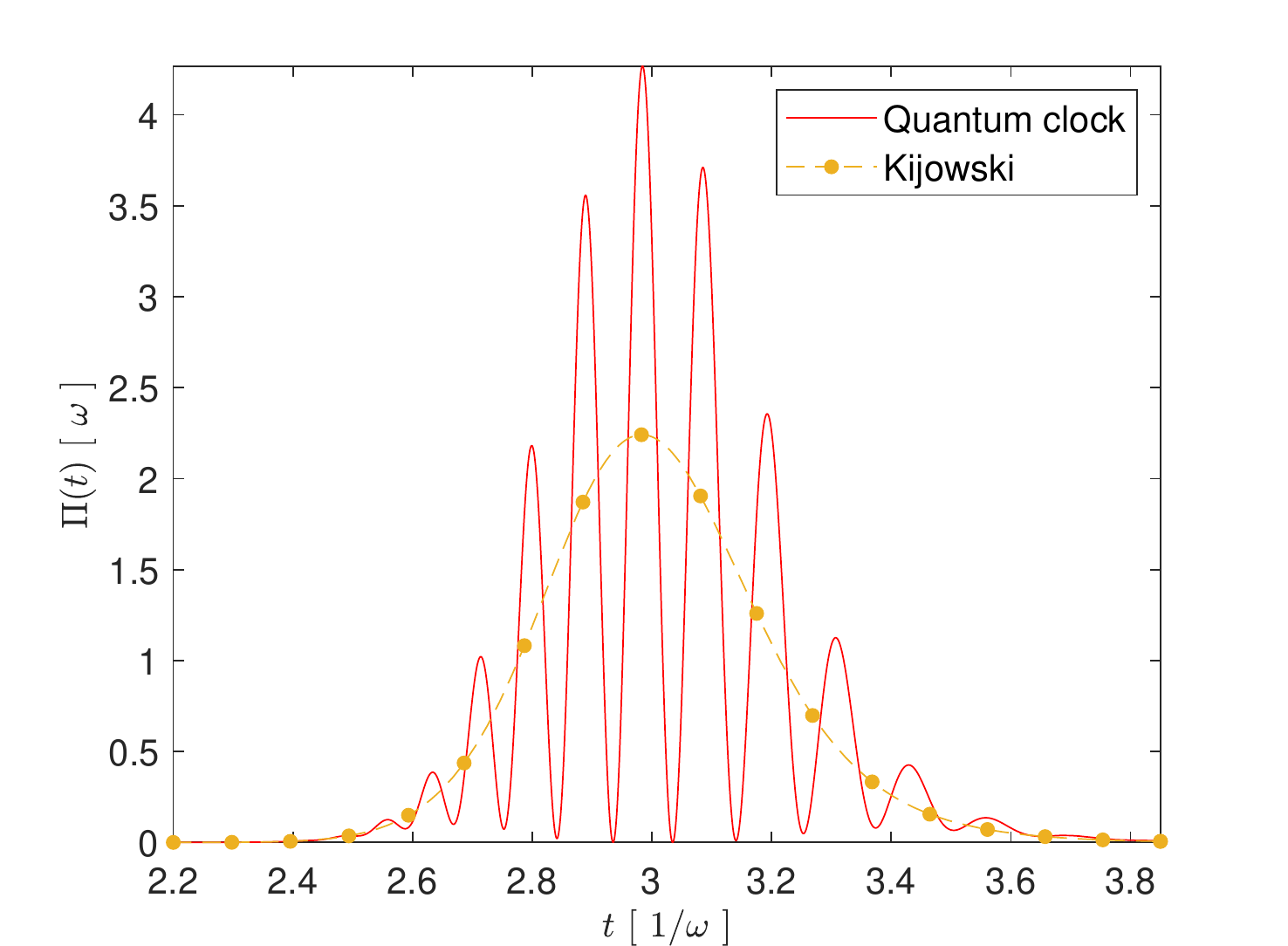}
    \caption{Comparison between $\Pi_{C}$ and $\Pi_{K}$ for $n=2$ counter-propagating packets that meet at the detector with the same average TOA. The parameters of each packet are $x_0 = -30 \ l_0$, $p_0 = 10 \ \hbar/l_0$, $x_1 = 45 \ l_0$, $p_1 = -15 \ \hbar/l_0$ and $\sigma_0 = \sigma_1 = 1 \ l_0$. The Kijowski's proposal lacks the interference fringes, while the quantum flux cannot be applied in this regime\label{fig:colliding}.}
\end{figure}%

\subsubsection{Counter-propagating packets}
We now apply the same procedure to test the combined Kijowski distribution of \cref{eq:TotalKijowskiSum}, using a superposition of packets supported for both positive and negative momenta values. To this extent, we consider a couple of counter-propagating Gaussian wave packets that cross the detector from two opposite directions with the same average TOA.  

In \cref{fig:colliding} we compare the TOA distributions $\Pi_{C}$ and $\Pi_{K}$, while $\Pi_{F}$ gives meaningless results in this non-classical regime. In this case, the anomalous behaviour of the quantum flux is not related to the presence of quantum backflow, it rather arises from our choice of packets with opposite momenta signs. In contrast to quantum clock, the Kijowski's proposal shows no interference pattern in this regime. This is due to the independent treatment of positive and negative momenta contributions in \cref{eq:Kijowski}, which are recombined by separately adding the two distributions $\Pi_{+}$ and $\Pi_{-}$.

For two counter-propagating packets, the quantum clock interference pattern is exactly the same as the one obtained in \cref{fig:overtaking}, where the packets approach the detector from the same side. This is due to $\Pi_{C}$ being proportional to $|\Psi(0,t)|^2$, which for two Gaussian wave packets gives
\begin{equation}
	\Pi_{C}(t) = \frac{1}{N_C} \left|\sum_{k=0}^{1}\frac{e^{-\frac{1}{2}(x_k^2 - 2ix_kp_k + itp_k^2)/(\sigma_k + it/\sigma_k)}}{\sqrt{\sigma_k + i t / \sigma_k}} \right|^2 \ , \
	\label{eq:QC_Symmetry}
\end{equation}
with $x_k$ and $p_k$ the initial average position and momentum of each packet. The substitutions $x_1 \to -x_1$ and $p_1 \to -p_1$ preserve the right-hand side of \cref{eq:QC_Symmetry} while translating the counter-propagating regime into the overtaking one (or the other way round). This has no effect on the quantum clock TOA which then predicts these regimes to be equivalent, when the detector is placed in the origin. This symmetry is not present in $\Pi_{K}$ and cannot be observed in $\Pi_{F}$, which is not well defined under such transformations.

\section{Experimental considerations\label{sec:SectionIII}}
In this section we analyse the possibility to observe the TOA interference patterns described above. 
\begin{figure*}
	\centering
	\subfloat[\label{fig:bec1}]{%
    		\includegraphics[width = \columnwidth]{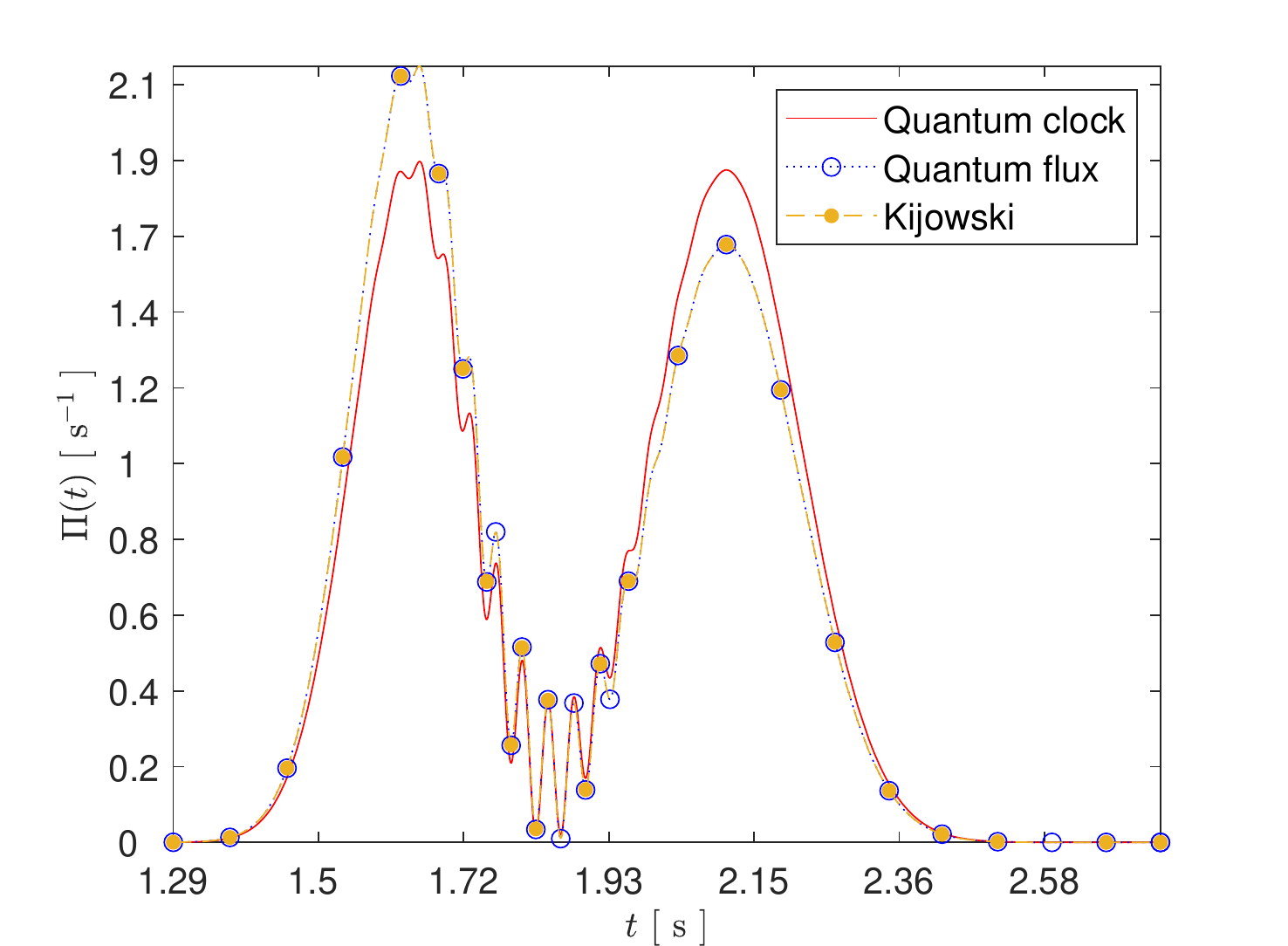}}%
   	\subfloat[\label{fig:bec2}]{%
    		\includegraphics[width = \columnwidth]{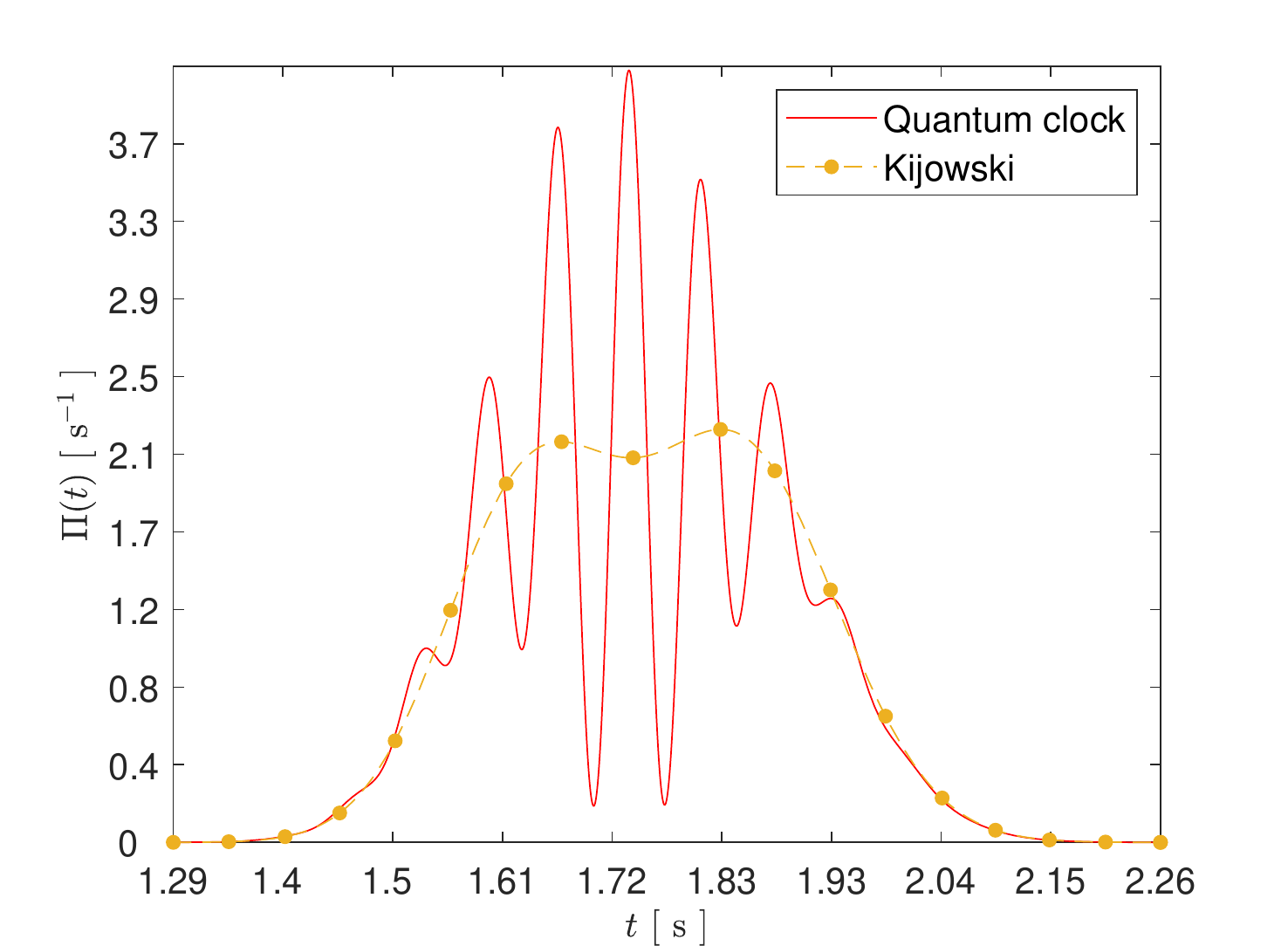}}
    \caption{Comparison between $\Pi_{C}$, $\Pi_{F}$ and $\Pi_{K}$ for a superposition of $n = 2$ Gaussian wave packets with periodic boundary conditions on a ring, which mimics the properties of a BEC split using Bragg diffraction. The TOA distributions are normalized in their respective plot interval. (a) The detector is located at $x=0$, the two packets are placed at $\bar{x}_0 = \bar{x}_1 = \SI{-1.39}{\milli \meter}$ and propagate clockwise with $\bar{p}_0 = \SI{1.07 e -25}{\milli \meter . \kilogram . \second^{-2}}$ and $\bar{p}_1 = \SI{0.83 e-25}{\milli \meter . \kilogram . \second^{-2}}$. Under these initial conditions, the packets partially overlap at the detector. For this reason, the effects of quantum backflow in $\Pi_{F}$ cannot be appreciated in this regime. (b) The second packet is initially placed at $\bar{x}_1 = \SI{-1.39}{\milli \meter}$, but it propagates counter-clockwise with $\bar{p}_1 = \SI{- 0.95 e-25}{\milli \meter . \kilogram . \second^{-2}}$. The Kijowski's TOA distribution shows no interference fringes, while the quantum flux cannot be applied in this non-definite momentum sign regime. Both (a) and (b) requires a temporal resolution of $\SI{0.1}{\second}$, which approximately corresponds to a detector size of $\SI{50}{\micro \meter}$.}
\end{figure*}

We consider the experimental parameters of a Bose-Einstein condensate (BEC) trapped in an accelerator ring \cite{art:VonKlitzing}. We prepare the BEC by using a cloud of $^{87}\text{Rb}$ subject to a harmonic potential. After releasing the trap, we load and accelerate the condensate along a ring of radius $R = \SI{443}{\micro\meter}$. We describe each atom of the cloud using a Gaussian wave packet that propagates on the circle of length $d = 2\pi R$, with periodic boundary conditions in the interval $[-d/2,d/2]$, thus both its momentum and energy take discrete eigenvalues $p_n = 2 \pi n\hbar/d$ and $E_n = 2 \pi^2 n^2 \hbar^2 / m d^2$. We focus only on the longitudinal degree of freedom, assuming that the cloud is sufficiently confined and that its transversal spreading is negligible with respect to length of the ring.\footnote{As long as the confinement is sufficiently strong, the atoms occupy only the lowest energy level of the transverse degree of freedom. In this case, the BEC truly behaves like the one-dimensional system described by \cref{eq:ConfinedBECWavePacket}.} Then, each wave packet is described by
\begin{equation}
	\psi(x,t) = \sum_{n=-N}^{N} a_n  e^{i p_n (x-\bar{x}_0)/\hbar} e^{-iE_nt/\hbar}  \ ,
	\label{eq:ConfinedBECWavePacket}
\end{equation}
where $x$ labels the curvilinear coordinate on the ring. For a packet with initial spatial width $\sigma_0 \ll d$, the wave plane coefficients read
\begin{equation}
	a_n = \left( \frac{4 \pi \sigma_0^2}{d^4}\right)^{\frac{1}{4}} e^{-(p_n - \bar{p}_0)^2 \sigma_0^2 / 2 \hbar^2} \ .
	\label{eq:Fourier_wave}
\end{equation}
Here $\bar{x}_0$ and $\bar{p}_0$ respectively denote the initial average position and momentum of the packet in the period $[-d/2,d/2]$.\footnote{For $\sigma_0 \ll d$ the tails of the Gaussian are negligible outside the period, so we can approximate the wave plane coefficients by integrating the packet on the whole space interval, i.e. by substituting the discrete momentum eigenvalues in the Fourier transform of the packet.}

We prepare the superposition of propagating packets by letting the BEC interact with a standing light wave. Due to Bragg diffraction, this coherently splits each packet with a momentum transfer that depends on the angle of incidence of light  \citep{art:BraggScattering}. Detection is then indirectly achieved, targeting the BEC with a focused laser beam (which is kept on for the entire duration of the experiment). The interaction with the beam extracts some atoms out of the condensate (with almost negligible feedback on the BEC). By collecting these atoms (or the photons scattered during the process), we can use the corresponding extraction rate to reconstruct the TOA distribution. In this case, the arrival of the particle is conditioned at the longitudinal position where the focused laser beam intersects the BEC. This corresponds to virtually place a detector in the ring, with temporal resolution controlled by the beam waist of the laser.

With the detector placed in the origin, we consider a packet with $\bar{x}_0 = - d/2$, $\sigma_0 = \SI{100}{\micro \meter}$ and $\bar{p}_0 = p_{450}$. Using Bragg diffraction, we coherently split it into another one with same initial position and width, but momentum $\bar{p}_1 = p_{350}$. In \cref{fig:bec1} we plot the TOA distributions given by the superposition of these two packets. With these initial conditions a partial overlap occurs, in a similar way it does in \cref{fig:overlapping}. In this regime $\Pi_{C}$ produces significantly different results from $\Pi_{F}$ and $\Pi_{K}$. The contributions from negative momentum components are negligible, hence the differences between $\Pi_C$ and $\Pi_K$ cannot be attributed to the incoherent sum in Kijowski's case, \cref{eq:TotalKijowskiSum}, where $\Pi_{-} \simeq 0$.

In \cref{fig:bec2} we perform the same analysis for two counter-propagating packets with initial positions $\bar{x}_0 = \bar{x}_1 = -d/2$, and momenta $\bar{p}_0 = p_{450}$, $\bar{p}_1 = -p_{400}$ with same initial standard deviation. Similarly to what happened in \cref{fig:colliding}, Kijowski's predicts no interference fringes with respect to the quantum clock proposal, so also in this regime an experimental comparison is desirable.

Taking into account experimental errors, we now estimate the minimum amount of data to achieve the results of this section. Let $\Pi_1$ and $\Pi_2$ be two TOA distributions. Consider $N_{s}$ detector clicks temporally distributed in a histogram with a certain number of bins. In the $k$-th bin there are two possible outcomes, which are described by a Bernoulli distribution: the click falls within the bin $\Delta t_k$ with probability $f_k$, or it does not with probability $1 - f_k$. Discrimination is possible whenever the error bar in the $k$-th bin is sufficiently smaller than the vertical separation between $\Pi_1$ and $\Pi_2$
\begin{equation}
	D = \left|\int_{\Delta t_k} dt\left[\Pi_1(t)-\Pi_2(t)\right]\right| \ , 
\end{equation}
namely when $\varepsilon_k < D/2$, with $\varepsilon_k$ the standard error on the $k$-th bin. For a Bernoulli distribution $\varepsilon_k = \sqrt{f_k(1-f_k)/N_s}$, which implies that the condition to achieve experimental discrimination becomes
\begin{equation}
	N_{s} > \frac{4f_k}{D^2}\left(1-f_k\right) \ .
	\label{eq:LowerBoundSample}
\end{equation}
Here $f_k$ is approximated by the experimental sample in terms of the relative frequencies $N_k/N_{s}$, with $N_k$ the number of clicks in the $k$-th bin. Then, through a conventional $\chi^2$ test we can compare the experimental data with the different theoretical predictions to determine which of the TOA proposals correctly reproduces the measurement outcomes.

In the case of \cref{fig:bec1} we choose as bin $ \Delta t_k = [\SI{1.29}{\second}, \SI{1.83}{\second}]$. We theoretically estimate $f_k$ by integrating the TOA distribution that maximises the lower bound of \cref{eq:LowerBoundSample} in $\Delta t_k$. In this regime $N_{s} \simeq 300$ clicks (i.e. single packet detections) represent a sufficient amount of data to discriminate between $\Pi_{C}$ and $\Pi_{K}$.

\section*{Discussion and perspectives}
In this paper we reviewed some of the proposals that aim to solve the quantum TOA problem. We compared the Kijowski's, the quantum flux and the quantum clock TOA distributions, using different superpositions of Gaussian wave packets in multiple regimes. Although inequivalent, the Kijowski's and the quantum flux distributions behave in a similar way, except when the quantum backflow contributions are not negligible. Instead, the quantum clock TOA produces significantly different results, in particular when the two packets are counter-propagating. In this regime the Kijowski's distribution predicts no interference fringes, while the quantum clock yields exactly the same pattern that occurs for two right-moving overtaking packets. This symmetry is not reproduced by the Kijowski's proposal, while it cannot be observed in the quantum flux.  

Such differences can be tested in the laboratory. In the last section, we discussed this possibility for an experimental implementation that combines BEC interferometry with Bragg diffraction. This design is capable of experimentally discriminating the predictions of Kijowski and of the quantum flux from those of the quantum clock approach. It also provides a useful test for the quantum clock symmetry, by comparing the overtaking and the counter-propagating regimes for two packets with the same absolute value of initial position and momentum.

\begin{acknowledgments}
L.M. and S.R. acknowledge support from MIUR Dipartimenti di Eccellenza 2018-2022, Project No. F11I18000680001, from EU H2020 QuantERA ERA-NET Cofund in Quantum Technologies, Quantum Information and Communication with High-dimensional Encoding (QuICHE), Grant Agreements 731473 and 101017733, from the U.S. Department of Energy, Office of Science, National Quantum Information Science Research Centers, Superconducting Quantum Materials and Systems Center (SQMS), Contract No. DE-AC02-07CH11359, and from the National Research Centre for HPC, Big Data and Quantum Computing (ICSC). This research was also funded by the National Science Centre, Poland, Project No. 2021/42/A/ST2/00017 (K.S.). For the purpose of Open Access, the authors have applied a CC-BY public copyright licence to any Author Accepted Manuscript (AAM) version arising from this submission.
\end{acknowledgments}

\bibliography{refs.bib}
\end{document}